\documentclass[a4paper,prd,twocolumn,aps,preprintnumbers,showpacs]{revtex4}
\usepackage{amssymb}
\usepackage{epsfig}
\usepackage{graphicx}% Include figure files
\usepackage{dcolumn}% Align table columns on decimal point
\usepackage{bm}% bold math
\usepackage{hyperref}
 
\def \be  {\begin{equation}}
\def \ee  {\end{equation}}
\def \beq  {\begin{equation}}
\def \eeq {\end{equation}}
\def \ba  {\begin{eqnarray}}
\def \ea  {\end{eqnarray}}
\def \baa {\begin{eqnarray*}}
\def \eaa {\end{eqnarray*}}
\def \bb  {\begin{thebibliography}}
\def \eb  {\end{thebibliography}}

\def \nn {\nonumber}

\def \lab #1 {\label{#1}}

\newcommand\as{\alpha_s}

\def \fracs #1#2 {\mbox{\small $\frac{#1}{#2}$}}

\def \bin #1#2 {{\left({#1}\atop{#2}\right)}}
\def\lapproxeq{{\ \lower 0.6ex \hbox{$\buildrel<\over\sim$}\ }}
\def\gapproxeq{{\ \lower 0.6ex \hbox{$\buildrel>\over\sim$}\ }}

\def\hepph  #1 {{hep-ph/#1 }}

\begin{document}
%
%\preprint{XXXX}

\title{QCD resummation for semi-inclusive hadron production processes}
\author{Daniele P. Anderle, Felix Ringer, Werner Vogelsang}
\affiliation{Institute for Theoretical Physics, 
T\"ubingen University, Auf der Morgenstelle 14, 72076 T\"ubingen, Germany}
%%%%%%%%%%%%%%%%
\begin{abstract}
We investigate the resummation of large logarithmic perturbative corrections 
to hadron production in electron-positron annihilation and semi-inclusive
deep-inelastic scattering. We find modest, but significant, enhancements of hadron 
multiplicities in the kinematic regimes accessible in present high-precision experiments. 
Our results are therefore relevant for the determination of hadron fragmentation functions
from data for these processes.
\end{abstract}

\date{\today}
\pacs{12.38.Bx, 13.85.Ni, 13.88.+e}
\maketitle

%%%%%%%%%%%%%%%%%%%%%%
\section{Introduction \label{intro}}
%%%%%%%%%%%%%%%%%%%%%%

Processes with identified final-state hadrons play important roles in QCD. Foremost, 
they provide crucial information on fragmentation functions and hence, ultimately, 
the hadronization mechanism. Modern analyses~\cite{defloriandss,strat,akk,hkns} of 
fragmentation functions variously use data for single-inclusive annihilation (SIA) 
$e^+e^-\to hX$, semi-inclusive deep-inelastic scattering (SIDIS),  $\ell p\to \ell hX$, 
and $pp\to hX$, where  $h$ denotes a final-state hadron. Hadron production
observables also serve as powerful probes of nucleon or nuclear structure. 
In particular, SIDIS measurements with polarized beams and/or targets 
have by now become indispensable tools for investigations of the spin structure
of the nucleon in terms of helicity parton distribution functions and transverse-momentum
dependent distributions~\cite{Nowak}. Finally, hadron production
data also test some of our key concepts in the theoretical analysis of QCD at
high energies, among them factorization, universality, and perturbative calculations.

In the present paper, we address higher-order perturbative corrections to two of the key 
hadron production processes, SIA $e^+e^-\to hX$ and SIDIS $\ell p\to \ell hX$. Our
study is very much motivated by the recent advent of data for these reactions
with unprecedented high precision. The BELLE collaboration at KEK has 
presented preliminary data~\cite{belle} for pion and kaon multiplicities in SIA that
extend over a wide range in values of the fragmentation variable $x_E=2E^h/\sqrt{s}$, 
where $E^h$ is the energy of the produced hadron in the $e^+e^-$ center-of-mass system, 
and $\sqrt{s}=10.52$~GeV the collision energy. The BELLE data cover the region
$0.2\leq x_E\leq 0.97$, with a very fine binning and extremely high precision partly at
the sub-1\% level. New preliminary high-statistics SIDIS data have  
been shown by the HERMES~\cite{hermes} and COMPASS~\cite{compass} 
collaborations over the past year or so. 

In the kinematic regimes accessed by these experiments, perturbative-QCD corrections 
are expected to be fairly significant. In case of $e^+e^-\to hX$ at BELLE,
as $x_E$ increases toward unity, the phase space for real-gluon radiation is very restricted, 
since most of the initial leptonic energy is used to produce the observed hadron and
a recoiling unobserved final state. When this happens, the infrared cancellations
between virtual and real-emission diagrams leave behind large logarithmic higher-order 
corrections to the basic $e^+e^-\to q\bar{q}$ cross section. These logarithms
are very similar in nature to the ``threshold logarithms'' encountered in hadronic
scattering processes when the energy of initial partons is just large enough to 
produce a given final state. Near $x_E=1$, it then becomes necessary 
to take the large corrections into account to all orders in the strong coupling,
a technique known as threshold resummation. 
The SIDIS measurements, on the other hand, are characterized by two scaling variables, Bjorken-$x$
and a variable $z$ given by the energy of the produced hadron over the energy of the 
virtual photon in the target rest frame. The cross section is typically defined to be
differential in both. Large logarithmic corrections to the SIDIS cross section 
arise when either one of the corresponding partonic variables becomes large. The most
important effects arise when both are large, which is typically the case for the 
presently relevant fixed-target kinematics. As we shall discuss, in this case the logarithmic 
terms can be simultaneously resummed to all orders within the threshold resummation
framework. 

Previous work~\cite{Sterman:2006hu} has established a close correspondence between 
threshold resummation for the Drell-Yan process, double-inclusive annihilation 
$e^+e^-\to h_1 h_2X$, and a variant of SIDIS for which one considers the cross section 
differential in the {\it product} $x z$ of the two scaling variables mentioned above, rather 
than in each of them separately. For this set of observables, the structure of the threshold 
logarithms turns out to be identical, up to trivial differences in the hard scattering functions 
that multiply the logarithms. One therefore can derive the resummation for each of the processes 
in the same manner, using exponentiation of eikonal diagrams in color-singlet processes~\cite{webs,LSV}. 
This approach was termed {\it crossed threshold resummation} in~\cite{Sterman:2006hu} and
will be the framework for our present analysis of SIDIS.  

Resummation for $e^+e^-\to hX$ was addressed in detail in Refs.~\cite{Cacciari:2001cw,Blum,Vogt,Procura:2011aq}.
In~\cite{Cacciari:2001cw} the next-to-leading logarithm (NLL) expressions were presented,
while extensions to next-to-next-to-leading logarithm and even next-to-next-to-next-to-leading logarithm
were provided in Refs.~\cite{Blum} and~\cite{Vogt}, respectively. The global analysis of fragmentation 
functions of Ref.~\cite{akk} in fact includes NLL threshold resummation effects for $e^+e^-\to hX$ for 
the lower-energy data. They are found to improve the theoretical description and the quality of the fit to the 
data. Given the very high precision of the new BELLE data, a new phenomenological analysis of resummation
effects is now timely and will be presented in this work. For now, we will restrict ourselves to resummation
at NLL, which captures the main effects. As for SIDIS, an expression for the NLL resummed cross section
was stated in~\cite{Cacciari:2001cw}, which turns out to be analogous to the related expressions for the 
rapidity-differential Drell-Yan cross section in terms of double-Mellin moments (see also~\cite{Cat,Bozzi:2007pn}). 
Phenomenological studies of resummation effects in SIDIS have never really been presented in the
literature, except briefly in~\cite{Vogelsang:2005tc}. In the present paper, we derive the 
NLL resummed expression for SIDIS, making use of the techniques of~\cite{Sterman:2006hu}. 
We also present numerical results as relevant for comparisons to the recent SIDIS data. 

Sections~\ref{sidis} and~\ref{epemresu} collect our technical derivations. Since the resummation 
for SIDIS is the main new result, we mostly focus on this process and only briefly review 
the well-known results for $e^+e^-\to hX$ and inclusive deep-inelastic scattering (DIS). 
Section~\ref{Pheno} presents our phenomenological results. 

%%%%%%%%%%%%%%%%%%%%%%%%%%%%%%%%%%%%%%%%%
\section{Resummation for SIDIS multiplicities \label{sidis}}
%%%%%%%%%%%%%%%%%%%%%%%%%%%%%%%%%%%%%%%%%

We consider semi-inclusive deep-inelastic scattering, $\ell(k)p(P)\to \ell(k')h(P_h)X$, 
where we have indicated the momenta of the involved particles. The momentum $q$
of the highly virtual photon exchanged between the incoming electron and proton is
given by $q=k-k'$.  We define the usual variables
\ba
Q^2 &\equiv& -q^2=-(k-k')^2, \nn\\[2mm]
x &\equiv& \frac{Q^2}{2P\cdot q}, \nn\\[2mm]
y &\equiv &\frac{P\cdot q}{P\cdot k}, \nn\\[2mm]
z &\equiv& \frac{P\cdot P_h}{P\cdot q}.
\ea
We have $Q^2=x y s$, with $\sqrt{s}$ the $\ell p$ center-of-mass energy. 
In the current fragmentation region that we will consider here, 
the SIDIS cross section may be written 
as~\cite{altarelli,Nason:1993xx,fupe,graudenz,deFlorian:1997zj,roth}: 
\begin{eqnarray}
\label{eq:sidis}
\frac{d^3\sigma^h}{dx dydz} &=& 
\frac{4\, \pi\alpha^2}{Q^2} 
\left[ \frac{1+(1-y)^2}{2y} {\cal F}_T^h(x,z,Q^2)\right. \nn\\[2mm]
&&\hspace*{1.1cm}\left.+ \frac{1-y}{y} {\cal F}_L^h(x,z,Q^2) \right],
\end{eqnarray}
where $\alpha$ is the fine structure constant. ${\cal F}_T^h$ 
and ${\cal F}_L^h$ are the transverse and the longitudinal structure 
functions; they are related to the more customary structure functions 
$F_1^h$ and $F_L^h$ by ${\cal F}_T^h\equiv 2F_1^h$ and 
${\cal F}_L^h\equiv F_L^h/x$. 

SIDIS hadron multiplicities are defined by 
\beq\label{Rdef}
R^h_{\mathrm{SIDIS}}\equiv\frac{d^3\sigma^h/dx dydz}{d^2\sigma/dxdy},
\eeq
where $d^2\sigma/dxdy$ is the cross section for inclusive DIS, $\ell p\to \ell X$, given by 
\beq
\label{eq:dis}
\frac{d^2\sigma}{dx dy} =
\frac{4\, \pi\alpha^2}{Q^2} 
\left[ \frac{1+(1-y)^2}{2y} {\cal F}_T(x,Q^2)+ 
\frac{1-y}{y} {\cal F}_L(x,Q^2) \right].
\eeq
Here ${\cal F}_T\equiv 2F_1$ and ${\cal F}_L\equiv F_L/x$, with the 
standard inclusive structure functions $F_1,F_L$. Usually, the numerator 
and denominator of~(\ref{Rdef}) are averaged over suitable bins in $x$ and $y$. 
In order to investigate higher-order effects on SIDIS multiplicities, we have to 
consider QCD corrections to both the SIDIS and the inclusive DIS cross section. 
Since the latter has been treated very extensively in the literature, we will focus here
on $d^3\sigma^h/dx dydz$ and only briefly summarize some of the known results
for $d^2\sigma/dxdy$. 

\subsection{SIDIS cross section at next-to-leading order, and Mellin moments \label{sec2A}}

Using factorization, the transverse and
longitudinal structure functions ${\cal F}_T^h\equiv 2F_1^h$ and 
${\cal F}_L^h\equiv F_L^h/x$ in (\ref{eq:sidis}) are 
given by ($i=T,L$)
\begin{eqnarray}
\label{eq:f1sidis}
{\cal F}_i^h(x,z,Q^2) &=&\sum_{f,f'} 
\int_x^1 \frac{d\hat{x}}{\hat{x}}\int_z^1 \frac{d\hat{z}}{\hat{z}}  f \left(\frac{x}{\hat{x}},
\mu^2\right) D^h_{f'} \left(\frac{z}{\hat{z}},\mu^2\right)\,\nn\\[2mm]
&&\times \,{\cal{C}}^i_{f'f}
\left(\hat{x},\hat{z},\frac{Q^2}{\mu^2},\alpha_s(\mu^2)\right) , 
\end{eqnarray}
where $f(\xi,\mu^2)$ denotes the distribution of parton $f=q,\bar{q},g$ in the nucleon
at momentum fraction $\xi$ and scale $\mu$, while $D^h_{f'} \left(\zeta,\mu^2\right)$
is the corresponding fragmentation function for parton $f'$ going to the
observed hadron $h$. For simplicity, we have set all factorization and renormalization
scales equal and collectively denoted them by $\mu$. The hard-scattering coefficient 
functions ${\cal{C}}^i_{f'f}$ can be computed in perturbation theory:
\beq\label{Cpert}
{\cal{C}}^i_{f'f}\,=\,C^{i,(0)}_{f'f}+\frac{\alpha_s(\mu^2)}{2\pi}C^{i,(1)}_{f'f}+
{\cal O}(\alpha_s^2),
\eeq
where, again, $i=T,L$. To lowest order (LO), only the process $\gamma^*q\to q$
contributes, and we have 
\ba\label{LO}
C^{T,(0)}_{qq}(\hat{x},\hat{z})&=& e_q^2\,
\delta(1-\hat{x})\delta(1-\hat{z}),\nn\\[2mm]
C^{L,(0)}_{qq}(\hat{x},\hat{z})&=&0,
\ea
with the quark's fractional charge $e_q$. Beyond LO, also gluons in the
initial or final state contribute. The full set of the first-order coefficient 
functions~\cite{altarelli,Nason:1993xx,fupe,graudenz,deFlorian:1997zj,roth}
$C^{i,(1)}_{f'f}$ are collected in the Appendix. 

Since threshold resummation can be derived in Mellin-moment space, it is 
useful to take Mellin moments of the structure functions ${\cal F}_T^h$ and ${\cal F}_L^h$.
Since $x$ and $z$ are independent variables, we take moments separately in 
both~\cite{altarelli,Stratmann:2001pb}. We define
\beq\label{moms}
\tilde{{\cal F}}^h_i(N,M,Q^2)\equiv\int_0^1 dx x^{N-1}\int_0^1 dz z^{M-1}\,
{\cal F}^h_i(x,z,Q^2).
\eeq
We then readily find from~(\ref{eq:f1sidis})
\ba\label{Melmo}
\tilde{{\cal F}}^h_i(N,M,Q^2)&=&\sum_{f,f'}\tilde{f}^N(\mu^2)\tilde{D}_{f'}^{h,M}(\mu^2)\nn\\[2mm]
&&\times\tilde{{\cal C}}^i_{f'f}\left(N,M,\frac{Q^2}{\mu^2},\alpha_s(\mu^2)\right),
\ea 
where
\ba
&&\tilde{f}^N(\mu^2)\equiv\int_0^1 dx x^{N-1}f(x,\mu^2),\nn\\[2mm]
&&\tilde{D}_{f'}^{h,M}(\mu^2)\equiv\int_0^1 dz z^{M-1}D^h_{f'}(z,\mu^2),\nn\\[2mm]
&&\tilde{{\cal C}}^i_{f'f}\left(N,M,\frac{Q^2}{\mu^2},\alpha_s(\mu^2)\right)\nn\\[2mm]
&&\hspace*{5mm}\equiv
\int_0^1 d\hat{x}\hat{x}^{N-1}\int_0^1 d\hat{z}\hat{z}^{M-1}\,
{\cal{C}}^i_{f'f}
\left(\hat{x},\hat{z},\frac{Q^2}{\mu^2},\alpha_s(\mu^2)\right) .\nn\\[2mm]
\ea
Thus, the Mellin moments of the structure functions are obtained from ordinary 
products of the moments of the parton distribution functions and fragmentation
functions, and double-Mellin moments of  the partonic hard-scattering functions.
For the perturbative expansion given in~(\ref{Cpert}), we have for the latter in
lowest oder according to~(\ref{LO}):
\ba
\tilde{C}^{T,(0)}_{qq}(N,M)&=&e_q^2 ,\nn\\[2mm]
\tilde{C}^{L,(0)}_{qq}(N,M)&=&0.
\ea
The corresponding moments of the next-to-leading order (NLO) terms $C^{i,(1)}_{f'f}$~\cite{Stratmann:2001pb} 
are also provided in the Appendix.

\subsection{Resummation of the SIDIS coefficient function}

As one can see from Eq.~(\ref{sidiseq8}), the NLO coefficient function 
$C_{qq}^{T,(1)}(\hat{x},\hat{z})$ receives large corrections near
$\hat{x},\hat{z}\to 1$. Choosing for simplicity $\mu_F=Q$, we have
\begin{eqnarray}
\label{sidiseq8res}
\nonumber
C_{qq}^{T,(1)}(\hat{x},\hat{z}) &\sim& e_q^2 C_F
\Bigg[ -8\delta(1-\hat{x})\delta(1-\hat{z})\\
\nonumber\\ \nonumber
&&\hspace*{-1.8cm}+2 \delta(1-\hat{x}) \left(\frac{\ln(1-\hat{z})}{1-\hat{z}}\right)_+ 
+2 \delta(1-\hat{z}) \left(\frac{\ln(1-\hat{x})}{1-\hat{x}}\right)_+  \\[2mm]
&&\hspace*{-1.8cm}+\frac{2}{(1-\hat{x})_+(1-\hat{z})_+}\Bigg],
\end{eqnarray}
corresponding in moment space to
\begin{eqnarray} 
\label{cqq1}
\tilde{C}_{qq}^{T,(1)}(N,M)&\sim&
e_q^2 C_F \Bigg[-8+\frac{\pi^2}{3}+\left(\ln\bar{N}+\ln\bar{M}\right)^2\Bigg],
\nn\\[2mm]
\end{eqnarray}
where $\bar{N}\equiv N{\mathrm{e}}^{\gamma_E}$, $\bar{M}\equiv 
M{\mathrm{e}}^{\gamma_E}$, with $\gamma_E$ the Euler constant. 
Here we have only kept contributions that are neither suppressed as $1/N$,
nor as $1/M$ in moment space. The terms given in~(\ref{sidiseq8res}) therefore 
always contain two distributions, one in $\hat{x}$ and one in $\hat{z}$. 

Threshold resummation addresses the logarithms in $\bar{N}$ and $\bar{M}$ 
to all orders in the strong coupling constant $\alpha_s$. More precisely,
it captures terms of the form $\alpha_s^k\ln^n N\ln^m M$, with $n+m\leq 2k$.
We now discuss the derivation 
of the resummed expression for the SIDIS coefficient function $\tilde{C}_{qq}^{T,(1)}
(N,M)$. Since the leading-order process is $\gamma^* q \to q$ scattering,
and since both the incoming and the outgoing quark are ``observed'',
the treatment has much in common with that for the total Drell-Yan cross section,
or for its ``crossed'' versions considered in Ref.~\cite{Sterman:2006hu}. 
A significant difference is, however, that in the present case two independent 
Mellin moments, $N$ and $M$, have to be considered. At large $N$ and $M$,
or equivalently $\hat{x}$ and $\hat{z}$, all gluon radiation from the basic process 
$\gamma^* q \to q$ becomes soft, since we have the relation~\cite{Sterman:2006hu}
\be
(1-\hat{x})+(1-\hat{z}) \approx\frac{2k^0}{Q},
\ee
where $k^0$ is the total energy of gluon radiation. 
The coefficient function may then be evaluated in the eikonal approximation for the
quarks and/or antiquarks involved in the hard scattering. In moment space, the eikonal 
hard scattering functions exponentiate, leading to~\cite{Sterman:2006hu,LSV,webs}
\begin{widetext}
\ba
\tilde{\cal{C}}^{T,{\mathrm{res}}}_{qq}(N,M,\alpha_s(Q^2)) &\propto&\exp \Bigg[
\int_0^{Q^2} {dk_\perp^2\over k_\perp^2}A_q\left(\as(k_\perp^2)\right) \Bigg\{
\int_{\frac{k_\perp^2}{Q^2}}^1\frac{d\xi}{\xi}
\left[ {\mathrm{e}}^{-N \xi-M\frac{k_\perp^2}{\xi Q^2}}-1
\right]+\ln \bar N+ \ln \bar M\Bigg\} \Bigg],\nn\\[2mm]
\label{nextA2}
\ea
\end{widetext}
which is valid to next-to-leading logarithmic (NLL) accuracy. Here, $A_q(\alpha_s)$ is 
a perturbative function:
\begin{equation}\label{exp_A}
 A_q(\alpha_s) = \frac{\alpha_s}{\pi} A_q^{(1)} + 
\left(\frac{\alpha_s}{\pi}\right)^2 A_q^{(2)}+ \dots,
\end{equation}
with 
\begin{equation}
 A_q^{(1)} = C_F, \quad A_q^{(2)} = \frac{1}{2} C_F\left[C_A
\left(\frac{67}{18}-\frac{\pi^2}{6}\right) - \frac{5}{9} N_f\right],
\end{equation}
where $C_F=4/3$, $C_A=3$ and $N_f$ is the number of active flavors.

Up to corrections that are exponentially suppressed at large $N,M$, the 
integral over $\xi$ in~(\ref{nextA2})  can be carried out analytically, and one finds
\ba
&&\hspace*{-9mm}
\int_{\frac{k_\perp^2}{Q^2}}^1\frac{d\xi}{\xi}
\left[ {\mathrm{e}}^{-N \xi-M\frac{k_\perp^2}{\xi Q^2}}-1
\right]+\ln \bar N+ \ln \bar M\nn\\[2mm]
&\approx&
2\left[ K_0 \left(\sqrt{NM}\,\frac{2k_{\perp}}{Q} \right)+
 \ln \left(\frac{k_{\perp}}{Q}\sqrt{\bar{N} \bar{M}}\right) \right],
 \label{Bessel}
\end{eqnarray}
where $K_0$ is a Bessel function. It arises when we extend the
$\xi$ integral to $0<\xi<\infty$. It is instructive to confront this
with the analogous expression for the Mellin-$N$ moments of the partonic
total Drell-Yan cross section, which reads~\cite{LSV,Sterman:2006hu}:
\ba
&&\hspace*{-9mm}
\int_{\frac{k_\perp^2}{Q^2}}^1\frac{d\xi}{\xi}
\left[ {\mathrm{e}}^{-N \left(\xi-\frac{k_\perp^2}{\xi Q^2}\right)}-1
\right]+2\ln \bar N\nn\\[2mm]
&\approx&
2\left[ K_0 \left(N\,\frac{2k_{\perp}}{Q} \right)+
\ln \left(\frac{k_{\perp}}{Q}\bar{N}\right) \right].
\end{eqnarray}
From this comparison one can immediately see that the result for
the resummed SIDIS cross section can be obtained from the one for the 
total Drell-Yan cross section by simply setting $\bar{N}\to\sqrt{\bar{N}\bar{M}}$.
In the case of SIDIS, the moments $N$ and $M$ independently fix the light-cone
plus component $\xi$ and the minus component $\zeta=k_\perp^2/(\xi Q^2)$ of 
the soft gluon momentum, resulting in the slightly more elaborate form of the 
exponent in~(\ref{Bessel}). Likewise, the $\overline{\mathrm{MS}}$-subtraction 
of collinear divergencies in the initial and the final state in SIDIS yields the terms 
$\ln\bar{N}$ and $\ln\bar{M}$, respectively, whereas in Drell-Yan one has the 
contribution $2\ln \bar N$ from the two initial partons.

With~(\ref{Bessel}), the final resummed coefficient function becomes in the 
$\overline{\mathrm{MS}}$ scheme:
\begin{widetext}
\beq\label{resummed3}
\tilde{\cal{C}}^{T,{\mathrm{res}}}_{qq}(N,M,\alpha_s(Q^2))=e_q^2 
H_{qq}\left(\alpha_s(Q^2),\frac{Q^2}{\mu^2}\right) 
\exp\left[2 \int_0^{Q^2} {dk_\perp^2\over k_\perp^2}A_q\left(\as(k_\perp^2)\right)
\left\{ K_0 \left(\sqrt{NM}\,\frac{2k_{\perp}}{Q} \right)+
 \ln \left(\frac{k_{\perp}}{Q}\sqrt{\bar{N} \bar{M}}\right) \right\}\right].
\eeq
\end{widetext}
Here we have included a perturbative function $H_{qq}$ that collects the
hard virtual corrections to $\gamma^* q \to q$ scattering. For resummation at NLL, one
needs to know $H_{qq}$ to first order in the strong coupling, which 
may be derived by expanding~(\ref{resummed3}) to ${\cal O}(\alpha_s)$ (keeping only
logarithmic terms in the exponent) and comparing to the explicit NLO 
expression~(\ref{cqq1}) for large $N,M$. One finds:
\be
H_{qq}\left(\alpha_s,\frac{Q^2}{\mu^2}\right)=1+\frac{\alpha_s}{2\pi}C_F
\left( -8 + \frac{\pi^2}{3}+3\ln \frac{Q^2}{\mu^2}\right) +{\cal O}(\alpha_s^2)\;.
\ee

We note that an alternative, but equivalent, 
form of the resummed result is~\cite{Cat,Cacciari:2001cw}
\ba\label{resummed3old}
\tilde{\cal{C}}^{T,{\mathrm{res}}}_{qq}(N,M,\alpha_s(Q^2))&=&e_q^2 
H_{qq}\left(\alpha_s(Q^2),\frac{Q^2}{\mu^2}\right) \nn\\[2mm]
&&\hspace*{-3.25cm}\times \exp\left[ \int_0^1 dx \frac{\xi^N -1}{1-\xi}
\int_{Q^2}^{(1-\xi)Q^2} \frac{dk_\perp^2}{k_\perp^2} 
A_q(\alpha_s(k_\perp^2))\right.\nn\\[2mm]
&&\hspace*{-2.5cm}+\int_0^1 d\zeta \frac{\zeta^M -1}{1-\zeta}
\int_{Q^2}^{(1-\zeta) Q^2} \frac{dk_\perp^2}{k_\perp^2} 
A_q(\alpha_s(k_\perp^2))\nn\\[2mm]
&&\hspace*{-3.9cm}\left.+ 
\int_0^1 dx \frac{\xi^N-1}{1-\xi}\int_0^1 d\zeta \frac{\zeta^M-1}{1-\zeta}
A_q\left(\alpha_s(Q^2 (1-\xi)(1-\zeta))\right)\right].\nn\\[2mm]
\ea
This expression can be obtained from~(\ref{nextA2}) by first writing the integrand as
\ba\label{nextAA2}
{\mathrm{e}}^{-N \xi-M\frac{k_\perp^2}{\xi Q^2}}-1
&=&\left({\mathrm{e}}^{-N \xi}-1\right)\left({\mathrm{e}}^{-M\frac{k_\perp^2}
{\xi Q^2}}-1\right)\nn\\[2mm]
&&+{\mathrm{e}}^{-N \xi}-1+{\mathrm{e}}^{-M\frac{k_\perp^2}
{\xi Q^2}}-1.
\ea
Including the integrals over $\xi$ and $k_\perp$ and substituting $k_\perp^2=\zeta\xi Q^2$, 
the first term on the right-hand-side of~(\ref{nextAA2}) yields
\be\label{part1}
\int_0^1 \frac{d\xi}{\xi}\left({\mathrm{e}}^{-N \xi}-1\right)
\int_0^1 \frac{d\zeta}{\zeta}\left({\mathrm{e}}^{-M \zeta}-1\right)
A_q\left(\as(\zeta\xi Q^2)\right).
\ee
Next, we deal with the term containing $\left({\mathrm{e}}^{-N \xi}-1\right)$ in~(\ref{nextAA2}). 
Combining with the logarithm $\ln\bar{N}$ in~(\ref{nextA2}) we find, up to corrections suppressed 
as $1/N$:
\ba\label{part2}
&&\int_0^{Q^2} {dk_\perp^2\over k_\perp^2}A_q\left(\as(k_\perp^2)\right)\left\{
\int_{\frac{k_\perp^2}{Q^2}}^1\frac{d\xi}{\xi}
\left( {\mathrm{e}}^{-N \xi}-1
\right)+\ln \bar N\right\}\nn\\[2mm]
&&\approx\int_0^1 \frac{d\xi}{\xi}
\left( {\mathrm{e}}^{-N \xi}-1\right)\int_{Q^2}^{\xi Q^2}\frac{dk_\perp^2}{k_\perp^2}
A_q\left(\as(k_\perp^2)\right).
\ea
Likewise,
\ba\label{part3}
&&\hspace{-3mm}\int_0^{Q^2} {dk_\perp^2\over k_\perp^2}A_q\left(\as(k_\perp^2)\right)\left\{
\int_{\frac{k_\perp^2}{Q^2}}^1\frac{d\xi}{\xi}
\left({\mathrm{e}}^{-\frac{Mk_\perp^2}{\xi Q^2}}-1
\right)+\ln \bar M\right\}\nn\\[2mm]
&&\hspace{1.5mm}\approx\int_0^1 \frac{d\zeta}{\zeta}
\left( {\mathrm{e}}^{-M \zeta}-1\right)\int_{Q^2}^{\zeta Q^2}\frac{dk_\perp^2}{k_\perp^2}
A_q\left(\as(k_\perp^2)\right).
\ea
Adding the expressions in Eqs.~(\ref{part1}),(\ref{part2}),(\ref{part3}),
and changing $\xi\to 1-\xi$, $\zeta\to 1-\zeta$, we recover~(\ref{resummed3old}). 
Our expression~(\ref{resummed3}) is however simpler and makes the close connection to the 
resummed expression for the total Drell-Yan cross section more transparent.

\subsection{Expansion to NLL}

At small argument, the Bessel function $K_0$ behaves as 
$$
K_0(x)=-\ln \left(x{\mathrm{e}}^{\gamma_E}/2\right) + {\cal O}(x^2 \ln x)\; .
$$
Therefore, logarithmic behavior of the integral in~(\ref{resummed3}) in $N$ and $M$
occurs only when $k_\perp$ is bounded from below, and the $K_0$ function
effectively acts as step function. To NLL, one finds 
the condition $k_\perp\geq Q/\sqrt{\bar{N}\bar{M}}$~\cite{LSV}:
\ba\label{resummed4}
&&\hspace*{-11mm}\tilde{\cal{C}}^{T,{\mathrm{res}}}_{qq}(N,M,\alpha_s(Q^2))=e_q^2 
H_{qq}\left(\alpha_s(Q^2),\frac{Q^2}{\mu^2}\right) \nn\\[2mm]
&&\hspace*{-8mm}\times\exp\left[2 \int_{\frac{Q^2}{\bar{N}\bar{M}}}^{Q^2} 
{dk_\perp^2\over k_\perp^2}A_q\left(\as(k_\perp^2)\right)
 \ln \left(\frac{k_{\perp}}{Q}\sqrt{\bar{N} \bar{M}}\right) \right].
\ea
The explicit NLL expansion of the right-hand side
is straightforward after inserting the standard expression for the running strong
coupling and, anyway, the result can be directly obtained from
the well-known one~\cite{Catani:1996yz} for Drell-Yan, setting $\bar{N}
\to \sqrt{\bar{N}\bar{M}}$ there. In the exponent in~(\ref{resummed4}) we obtain
\ba\label{h1h2a}
&&\hspace*{-2mm}\int_{\frac{Q^2}{\bar{N}\bar{M}}}^{Q^2} 
{dk_\perp^2\over k_\perp^2}A_q\left(\as(k_\perp^2)\right)
 \ln \left(\frac{k_{\perp}}{Q}\sqrt{\bar{N} \bar{M}}\right)\nn\\[2mm]
 &&\approx
h_q^{(1)}\left(\frac{\lambda_{NM}}{2}\right)\,\frac{\lambda_{NM}}{2b_0\alpha_s(\mu^2)}
+h_q^{(2)}\left(\frac{\lambda_{NM}}{2}, \frac{Q^2}{\mu^2},\frac{Q^2}{\mu_F^2}
\right),\nn\\[2mm]
\ea
where
\ba\label{h1h2}
\lambda_{NM} &\equiv&b_0\alpha_s(\mu^2)\left(\log\bar{N}+\log\bar{M}\right),\nn\\[2mm]
h_q^{(1)}(\lambda) &=& \frac{A_q^{(1)}}{2 \pi b_0 \lambda}\left[2\lambda + 
(1- 2\lambda) \ln(1-2\lambda)\right],\nonumber \\[2mm]
h_q^{(2)}\left(\lambda, \frac{Q^2}{\mu^2},\frac{Q^2}{\mu_F^2}\right) &= & -
\frac{A_q^{(2)}}{2 \pi^2 b_0^2} \left[2\lambda + \ln (1-2\lambda)\right] \nn\\[2mm]
&&\hspace*{-2cm} +\frac{A_q^{(1)}b_1}{2\pi b_0^3}\left[2\lambda + 
\ln(1-2\lambda)+ \frac{1}{2} \ln^2(1-2\lambda)\right] \nonumber  \\[2mm]
& &\hspace*{-2cm} + \frac{A_q^{(1)}}{2\pi b_0} \left[2 \lambda + 
\ln(1-2\lambda)\right] \ln\frac{Q^2}{\mu^2}  - \frac{A_q^{(1)}}{\pi b_0} \lambda
 \ln \frac{Q^2}{\mu_F^2},\nn\\[2mm]
 \ea
with
\ba
b_0 &=& \frac{11 C_A - 4 T_R N_f}{12\pi},\nn\\[2mm]
b_1 &=&\frac{17 C_A^2 - 10 C_A T_R N_f -6 C_F T_R N_f}{24\pi^2}.
\ea
The functions $h_q^{(1)}$, $h_q^{(2)}$ collect all leading-logarithmic
and NLL terms in the exponent, which are of the form 
$\alpha_s^k \ln^n \bar{N} \ln^m \bar{M}$ with 
$n+m=k+1$ and $n+m=k$, respectively. Note that we have kept the factorization 
and renormalization scales arbitrary in the above expressions. 
The standard Drell-Yan result is recovered by setting $\lambda_{NM}/2\to \lambda_{\mathrm{DY}}$,
where $\lambda_{\mathrm{DY}}=b_0\alpha_s(\mu^2)\log\bar{N}$.

\subsection{Inverse Mellin transforms \label{tran}}
%%%%%%%%%%%%%%%%%%%%%%%%%%%%%%%%%%%%%%%%%%%%%%%%
\begin{figure*}[t!]
\centering
\includegraphics[width=16cm,clip]{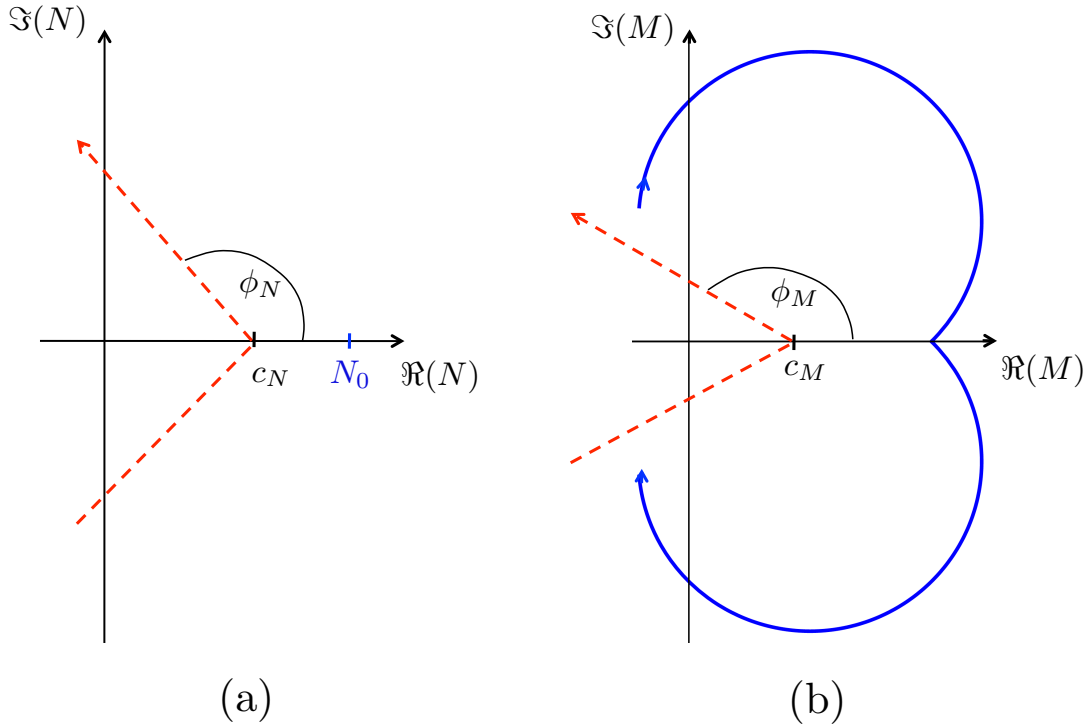}
\vspace*{-1.5cm}
\caption{\label{fig1}
\sf Contours for the inverse Mellin transforms, (a) for ${\cal C}_N$, (b) for ${\cal C}_M$.
We have defined $N_0=L_0/c_M$, with $L_0$ given in~(\ref{landau}).
In (b), the solid line depicts the location of the Landau pole as $N$ moves along a part of its 
contour (see text).}
\end{figure*}
%%%%%%%%%%%%%%%%%%%%%%%%%%%%%%%%%%%%%%%%%%%%%%%%

As the exponentiation of soft-gluon corrections is achieved in Mellin 
moment space, the hadronic structure function is obtained by taking the 
inverse Mellin transforms of Eq.~(\ref{moms}):
\begin{equation}\label{eq:inverse}
{\cal F}^h_i(x,z,Q^2)= \int_{{\cal C}_N}
\frac{d N}{2\pi i}x^{-N}  \int_{{\cal C}_M}
\frac{d M}{2\pi i}z^{-M}
\tilde{{\cal F}}^h_i(N,M,Q^2),
\end{equation}
where ${\cal C}_N$ and ${\cal C}_M$ denote integration contours in the complex
plane, one for each Mellin inverse. When performing an inverse Mellin transform, 
the contour usually has to be chosen in such a way that all singularities of the integrand 
lie to its left. However, as can be seen from Eq.~(\ref{h1h2a}),(\ref{h1h2}), the resummed 
cross section has a Landau singularity at $\lambda_{NM} = 1$ or 
\beq\label{landau}
N\,M={\mathrm{e}}^{1/(\alpha_s b_0)-2\gamma_E}\equiv L_0,
\eeq
as a result of the divergence of the running coupling $\alpha_s$ 
in Eq.~(\ref{resummed3}) for $k_\perp \to \Lambda_{\mathrm{QCD}}$. 
For the Mellin inversions, we adopt the \textit{minimal prescription} 
developed in Ref.~\cite{Catani:1996yz} to deal with the Landau pole. For 
this prescription the contours are chosen to lie to the {\it left} of the Landau 
singularity. In order to achieve this, we first choose the contour for the 
$N$-integration in the upper complex half plane in the standard way 
(see Fig.~\ref{fig1}(a)) as
\beq
N=c_N + z_N \,{\mathrm{e}}^{i\phi_N},
\eeq
where $c_N$ is a postive real constant, $\phi_N\sim 3\pi/4$ and $z_N\in[0,\infty]$. 
For the lower branch of the contour,  one simply uses the complex conjugate of $N$. 
The contour for the $M$-integration is parameterized in a similar fashion, with a
constant $c_M$, an integration parameter $z_M$ and an angle $\phi_M$ that
we will address shortly. The fact that both contours are tilted into the half-plane with 
negative real part improves the numerical convergence of the integration, since 
contributions with negative real part are exponentially suppressed by the factors
$x^{-N}$, $z^{-M}$ in Eq.~(\ref{eq:inverse}). 

In accordance with the minimal prescription~\cite{Catani:1996yz}, the parameter $c_N$ is chosen to be 
smaller than $N_0\equiv L_0/c_M$. As $N$ moves along its contour from a point with large
negative imaginary part to a point with large positive one, the Landau pole
given by Eq.~(\ref{landau}) describes a trajectory shown by the solid line 
in Fig.~\ref{fig1}(b). The angle $\phi_M$ is now chosen in such a way that
the $M$-integration contour (shown by the dashed line) avoids this trajectory. 
The larger the imaginary part of $N$, the larger the angle $\phi_M$ needs to become. 
Of course, ultimately as $\Im(N)\to\pm\infty$, the Landau pole moves to the origin
in the $M$-plane, and the contour falls onto the real $M$-axis. However, 
as described above, the contributions from such large values of $N$ are 
extremely suppressed. We note that a similar approach to the choice
of contours was discussed in Ref.~\cite{Sterman:2000pt}, where combined
inverse Mellin and Fourier transforms were considered. An alternative approach
is to expand the resummed formula to high perturbative orders. At each 
finite order,  the Landau pole is not present and, therefore, standard Mellin
contours can be chosen. 

We match the resummed cross section to 
the NLO one by subtracting the $O(\alpha_s)$ expansion of the resummed 
expression and adding the full NLO cross section. 
This ``matched'' cross section consequently not only resums the large 
threshold logarithms to all orders, but also contains the full NLO 
results for the $q\to q$, $q\to g$ and $g\to q$ channels. We will occasionally 
also consider a resummed cross section that has not been matched to 
the NLO one. We will refer to such a cross section as ``unmatched''.

\subsection{Resummation for inclusive DIS}

In order to obtain resummed predictions for the SIDIS hadron multiplicities defined in~(\ref{Rdef}), 
we also need the resummation for the inclusive cross section. In this case, there is only one 
variable $x$, and standard Mellin-moment resummation techniques~\cite{Schaefer:2001uh,neubert} 
may be applied. The inclusive structure functions ${\cal F}_T$ and ${\cal F}_L$ introduced
in Eq.~(\ref{eq:dis}) can be written as 
\beq
\label{eq:f1dis}
{\cal F}_i(x,Q^2) =\sum_f 
\int_x^1 \frac{d\hat{x}}{\hat{x}} f \left(\frac{x}{\hat{x}},
\mu^2\right) {\cal{C}}^i_f
\left(\hat{x},\frac{Q^2}{\mu^2},\alpha_s(\mu^2)\right) .
\eeq
We refer the reader to Ref.~\cite{fupe} for the NLO expressions
for the coefficient functions ${\cal{C}}^i_f$. Resummation may
be performed by again introducing Mellin moments in $x$.
The resummed DIS coefficient function for the structure function ${\cal F}_T$ reads
to NLL in moment space:
\ba\label{resummed1}
\tilde{\cal{C}}^{T,{\mathrm{res}}}_q&=&e_q^2 H_q\left(\alpha_s(Q^2),\frac{Q^2}{\mu^2}
\right) \exp\left[ \int_0^1 d\xi \frac{\xi^N -1}{1-\xi}\right.\nn\\[2mm]
&&\hspace*{-1.5cm}\left.\times\left\{
\int_{Q^2}^{(1-\xi) Q^2} \frac{dk_\perp^2}{k_\perp^2} 
A_q(\alpha_s(k_\perp^2))+\frac{1}{2}B_q\left(\alpha_s((1-\xi)Q^2)\right)
\right\}\right],\nn\\[2mm]
\ea
where the function $A_q(\alpha_s)$ is as in~(\ref{exp_A}). The perturbative 
function $B_q(\alpha_s)$ is given by 
\begin{equation}
B_q(\alpha_s) = \frac{\alpha_s}{\pi} B_q^{(1)} + {\cal O}(\alpha_s^2), 
\end{equation}
with 
\begin{equation}
B_q^{(1)} = -\frac{3}{2} C_F.
\end{equation}
Finally, the hard-scattering coefficient reads 
\ba\label{Hq}
&& \hspace*{-2mm}
H_q\left(\alpha_s(Q^2),\frac{Q^2}{\mu^2}\right)=1+ \frac{\alpha_s}{2\pi} 
C_F \left(-\frac{9}{2} - \frac{\pi^2}{6} +
\frac{3}{2}\ln \frac{Q^2}{\mu^2}\right)\nn\\[2mm]
&& \hspace*{3.3cm}+{\cal O}(\alpha_s^2).
\ea

The exponential in~(\ref{resummed1}) can evidently be written as $\Delta_q^N\times J_q^N$,
where
\ba
\log \Delta_q^N&\equiv&\left[\int_0^1 dx \frac{\xi^N -1}{1-\xi}
\int_{Q^2}^{(1-\xi)^2 Q^2} \frac{dk_\perp^2}{k_\perp^2} 
A_q(\alpha_s(k_\perp^2))\right],\nn\\[2mm]
\log J_q^N&\equiv&\left[ \int_0^1 d\xi \frac{\xi^N -1}{1-\xi}\left\{
\int_{(1-\xi)^2Q^2}^{(1-\xi) Q^2} \frac{dk_\perp^2}{k_\perp^2} 
A_q(\alpha_s(k_\perp^2))\right.\right.\nn\\[2mm]
&&\hspace*{4mm}\left.+\frac{1}{2}B_q(\alpha_s((1-\xi)Q^2))\right\} \Bigg].
\ea
The NLL expansions of these functions can be performed in the same way
as described in the previous subsection. One obtains:
\begin{equation}\label{eq:exponent}
\log\Delta_q^N\,=\, \ln \bar N h_q^{(1)}(\lambda) 
+  h_q^{(2)} \left(\lambda, \frac{Q^2}{\mu^2},\frac{Q^2}{\mu_F^2}\right), 
\end{equation}
with the functions $h_q^{(1)},h_q^{(2)}$ given in~(\ref{h1h2}). 
Furthermore, 
\beq
\label{lnjfun}
\ln J_q^N=\ln \bar{N} f_q^{(1)}(\lambda) +f_q^{(2)}\left(\lambda, \frac{Q^2}{\mu^2}\right),
\eeq
with~\cite{Schaefer:2001uh}
\ba
\label{fllh}
f_q^{(1)}(\lambda) &=& h_q^{(1)}\left(\frac{\lambda}{2}\right)- h_q^{(1)}(\lambda), \nonumber \\[2mm]
f_q^{(2)}\left(\lambda,\frac{Q^2}{\mu^2}\right) &=& 
2 h_q^{(2)}\left(\frac{\lambda}{2}, \frac{Q^2}{\mu^2},1\right)- 
h_q^{(2)}\left(\lambda, \frac{Q^2}{\mu^2},1\right) \nn\\[2mm]
&+&\frac{B_q^{(1)}}{2\pi b_0}\ln(1-\lambda).
\ea

\section{Resummation for $e^+e^-\to hX$ \label{epemresu}}

Hadron multiplicities in $e^+e^-\to hX$ are defined by
\beq\label{Reedef}
R^h_{e^+e^-}\equiv\frac{1}{\sigma^{\mathrm{tot}}}\frac{d^2\sigma^h}{dx_E d\cos\theta},
\eeq
where $d^2\sigma^h/dx_E d\cos\theta$ is the differential cross section 
for the production of the hadron $h$ at angle $\theta$ relative to the initial positron.
Furthermore,
\beq
x_E\equiv\frac{2P_h\cdot q}{Q^2},
\eeq
where $P_h$ and $q$ are the momenta of the produced hadron and the intermediate virtual 
photon respectively and $Q^2=q^2$. $\sigma^{\mathrm{tot}}$ is the total cross section for 
$e^+e^-\to {\mathrm{hadrons}}$. To first order in $\alpha_s$ it reads:
\be
\sigma^{\mathrm{tot}}=\frac{4\pi\alpha^2}{3Q^2}N_C\sum_q e_q^2
\left(1+\frac{\alpha_s}{\pi}\right),
\ee
where $N_C=3$ is the number of colors.

As in the case of SIDIS, one can write the cross section $d^2\sigma^h/dx_E d\cos\theta$
in terms of structure functions~\cite{fupe,Nason:1993xx}:
\ba
\label{eq:epem}
\frac{d^2\sigma^h}{dx_E d\cos\theta} &=&
\frac{\pi\alpha^2}{Q^2} N_C \left[ \frac{1+\cos^2\theta}{2} 
\hat{{\cal F}}^h_T(x_E,Q^2)\right.\nn\\[2mm]
&&+ \sin^2\theta\, \hat{{\cal F}}^h_L(x_E,Q^2) \Big],
\ea
where 
\beq
\label{eq:f1epem}
\hat{{\cal F}}_i^h(x_E,Q^2) =\sum_f 
\int_{x_E}^1 \frac{d\hat{z}}{\hat{z}} D_f^h \left(\frac{x_E}{\hat{z}},
\mu^2\right) \hat{{\cal{C}}}^i_f
\left(\hat{z},\frac{Q^2}{\mu^2},\alpha_s(\mu^2)\right),
\eeq
with the fragmentation functions $D_f^h$ introduced in subsection~\ref{sec2A}.
To lowest order, only the partonic channel $e^+e^-\to qX$ contributes, for which
\ba\label{LOepem}
\hat{C}^{T,(0)}_q(\hat{z})&=& e_q^2\,\delta(1-\hat{z}),\nn\\[2mm]
\hat{C}^{L,(0)}_q(\hat{z})&=&0.
\ea
Again we refer the reader to the previous literature~\cite{altarelli,fupe,Kretzer:2000yf} 
for the NLO expressions for the coefficient functions. After taking Mellin
moments in $x_E$, the resummed result for the corresponding hard-scattering 
function turns out to be identical to that 
in~(\ref{resummed1})~\cite{Cacciari:2001cw,Blum,Vogt}, except for a
change $-\pi^2/6 \to 5 \pi^2/6$ in the coefficient $H_q$ in~(\ref{Hq}).

%%%%%%%%%%%%%%%%%%%%%%%%%%%%%%%%%%%%%%%%%%%%%%%%
\section{Phenomenological Results \label{Pheno}}
%%%%%%%%%%%%%%%%%%%%%%%%%%%%%%%%%%%%%%%%%%%%%%%%

We now investigate the numerical size of the threshold resummation effects for the two semi-inclusive hadron production 
processes discussed above, SIDIS and $e^+e^-\to hX$. We focus entirely on pion production in this work,  for which
the theory is expected to be under best control. We also consider a proton target throughout this work.
For the parton distribution functions we use the NLO ``Martin--Stirling--Thorne--Watt'' 
(MSTW 2008) set of~\cite{Martin:2009iq}, whereas we choose the NLO ``de Florian--Sassot--Stratmann'' 
(DSS)~\cite{defloriandss} pion 
fragmentation functions. In this set, fragmentation functions for charged pions $\pi^\pm$ are separately available. 
We note that the parton distributions and fragmentation functions are provided in $x$ (or $z$) space, whereas according to 
Eq.~(\ref{Melmo}) we need their Mellin moments. To obtain the latter, we first fit suitable functions of the form 
$Ax^\alpha(1-x)^\beta$ times a polynomial in $x$ to the distributions. It is then straightforward to take
Mellin moments of the fitted functions analytically and use them in the numerical code. We have checked that the
accuracy of the fit is overall very good.

\subsection{Results for SIDIS}

%%%%%%%%%%%%%%%%%%%%%%%%%%%%%%%%%%%%%%%%%%%%%%%%
\begin{figure}[t]
\vspace*{-2mm}
\hspace*{-3.8cm}
\includegraphics[width=0.55\textwidth,angle=90]{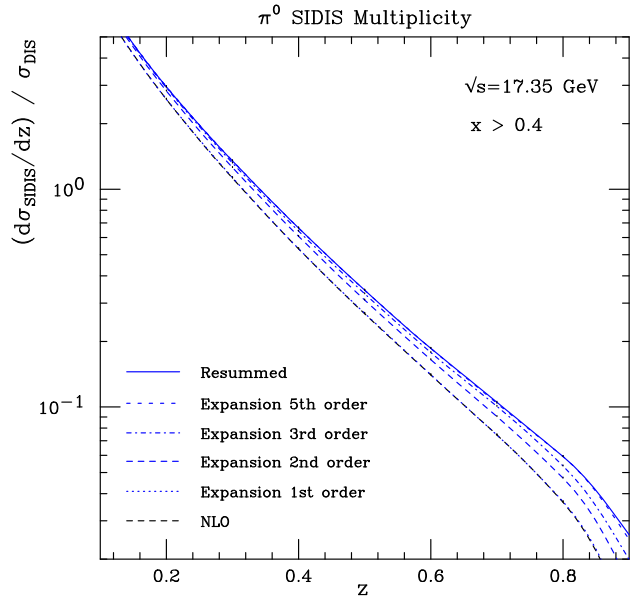}
\vspace*{-1.5cm}
\caption{\label{fig2}\sf SIDIS multiplicity for $\pi^0$. Kinematics are as for the COMPASS 
measurements~\cite{compass} (see text) with, however, an additional cut $x>0.4$ to 
enhance the contribution from the threshold regime. The lower two lines which are
almost indistinguishable show the NLO result and the first-order expansion of the 
resummed result (without matching). The top line shows the full NLL resummed result;
the lines in between display various fixed-order expansions of the latter.}
\end{figure}
%%%%%%%%%%%%%%%%%%%%%%%%%%%%%%%%%%%%%%%%%%%%%%%%
We start by examining the overall effects of threshold resummation for SIDIS, using the 
kinematics relevant for the COMPASS SIDIS measurements~\cite{compass} as an example. 
COMPASS uses a muon beam of energy 160 GeV incident on a proton fixed target. 
The resulting center-of-mass energy is $\sqrt{s}\approx$17.4 GeV. The kinematic cuts employed by 
COMPASS are $0.041<x<0.7$, $0.1<y<0.9$, $Q^2>1$ GeV${}^2$ and $W^2=Q^2(1-x)/x+
m_p^2>49$ GeV${}^2$, where $m_p$ is the proton mass. We choose the renormalization 
and factorization scales as $Q$ and consider the SIDIS multiplicity for neutral
pions $\pi^0$ as a function of $z$, but integrating numerator and denominator of Eq.~(\ref{Rdef}) over
$x$ and $y$. We observe that the range of $x$ probed by the full COMPASS data sample
extends down to fairly low values, where one could be quite far from the threshold regime. 
We therefore first consider a lower cut of $x>0.4$.  Figure~\ref{fig2} shows our results for
the NLO and the resummed multiplicities, along with those for expansions of the resummed
cross section to various finite orders in $\alpha_s$. In each case, the denominator of
the multiplicity, the inclusive-DIS cross section, has been treated in the same fashion
as the SIDIS one in the numerator. With the exception of the first-order
expansion, all beyond-NLO results have been matched to the NLO one (separately in the
numerator and the denominator) as described at the end of
Sec.\ref{tran}. We first of all note that the first-order expansion of the resummed cross 
section agrees very well with the full NLO one, which demonstrates that for the chosen 
kinematics the threshold regime strongly dominates the NLO cross section. The full resummed 
result shows a marked increase over the NLO one, in particular at high $z$, and the 
higher-order expansions converge nicely to the resummed result. Clearly, the $\alpha_s^2$ 
and $\alpha_s^3$ contributions generated by resummation are still significant. 

\begin{figure}[t]
\vspace*{-1cm}
\hspace*{-3.8cm}
\includegraphics[width=0.55\textwidth,angle=90]{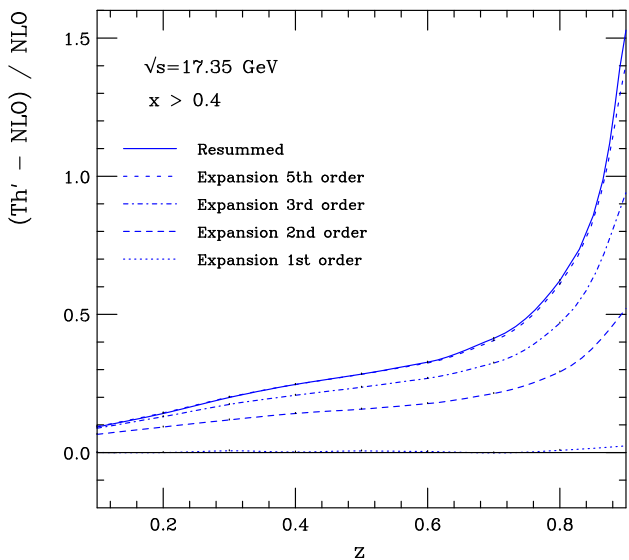}
\vspace*{-1.5cm}
\caption{\label{fig3}\sf Ratios $({\mathrm{Th' - NLO}})/{\mathrm{NLO}}$,
where ${\mathrm{Th'}}$ corresponds to the SIDIS multiplicity at higher
orders as generated by resummation.
The first-order expansion of the resummed cross section has
not been matched since it would otherwise be identical to NLO.}
\end{figure}

The results can also be studied as ratios $({\mathrm{Th' - NLO}})/{\mathrm{NLO}}$, where
${\mathrm{Th'}}$ denotes any of the higher-order SIDIS multiplicities generated by 
resummation. Figure~\ref{fig3} shows these ratios for the unmatched first-order expansion, 
the matched higher-order expansions and the full resummed result. The good agreement of 
NLO and the first-order expansion is evident, as are the large resummation effects at high $z$. 
%%%%%%%%%%%%%%%%%%%%%%%%%%%%%%%%%%%%%%%%%%%%%%%%
%%%%%%%%%%%%%%%%%%%%%%%%%%%%%%%%%%%%%%%%%%%%%%%%

\begin{figure}[t]
\vspace*{-9mm}
\hspace*{-3.8cm}
\includegraphics[width=0.55\textwidth,angle=90]{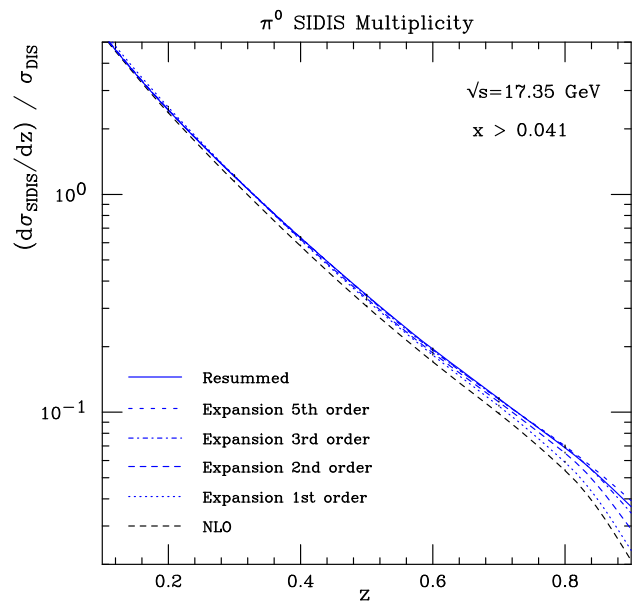}
\vspace*{-1.5cm}
\caption{\label{fig4}\sf Same as Fig.~\ref{fig2}, but for $x>0.041$.}
\end{figure}

\begin{figure}[t]
\vspace*{-1cm}
\hspace*{-3.8cm}
\includegraphics[width=0.55\textwidth,angle=90]{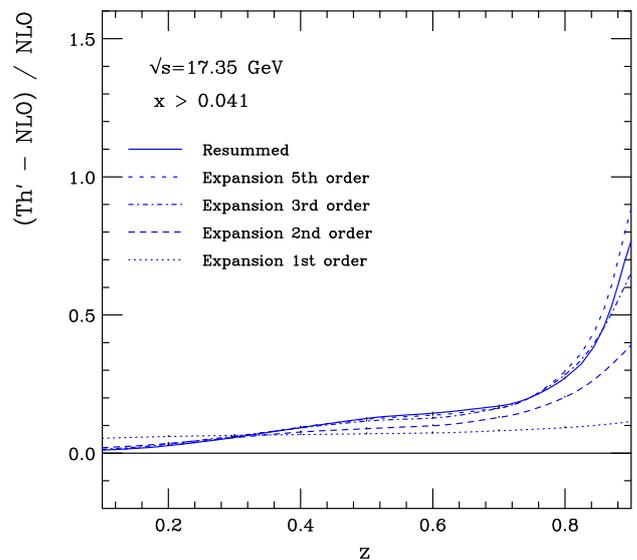}
\vspace*{-1.5cm}
\caption{\label{fig5}\sf Same as Fig.~\ref{fig3}, but for $x>0.041$.}
\end{figure}

We now extend the $x$-range to the full region $0.041<x<0.7$ covered by COMPASS.
Figs.~\ref{fig4} and~\ref{fig5} show the corresponding results, where all lines directly 
correspond to the ones shown in Figs.~\ref{fig2} and~\ref{fig3} for the case $x>0.4$.  
One can see that the resummation effects are generally smaller now, even though they
remain significant at high $z$. As expected, the agreement between NLO and the 
first-order expansion of the resummed cross section is worse now, but it typically
remains at the 10\% level or better. The second-order expansion captures most
of the full resummation effects; the yet higher orders converge somewhat more slowly now
to the resummed result. Figure~\ref{fig5} shows the corresponding ratios 
$({\mathrm{Th' - NLO}})/{\mathrm{NLO}}$, where again ${\mathrm{Th'}}$ denotes 
any of the SIDIS multiplicities computed at higher orders.
%%%%%%%%%%%%%%%%%%%%%%%%%%%%%%%%%%%%%%%%%%%%%%%%
%%%%%%%%%%%%%%%%%%%%%%%%%%%%%%%%%%%%%%%%%%%%%%%%
%%%%%%%%%%%%%%%%%%%%%%%%%%%%%%%%%%%%%%%%%%%%%%%%
%%%%%%%%%%%%%%%%%%%%%%%%%%%%%%%%%%%%%%%%%%%%%%%%

\begin{figure}[t]
\vspace*{-9mm}
\hspace*{-3.8cm}
\includegraphics[width=0.55\textwidth,angle=90]{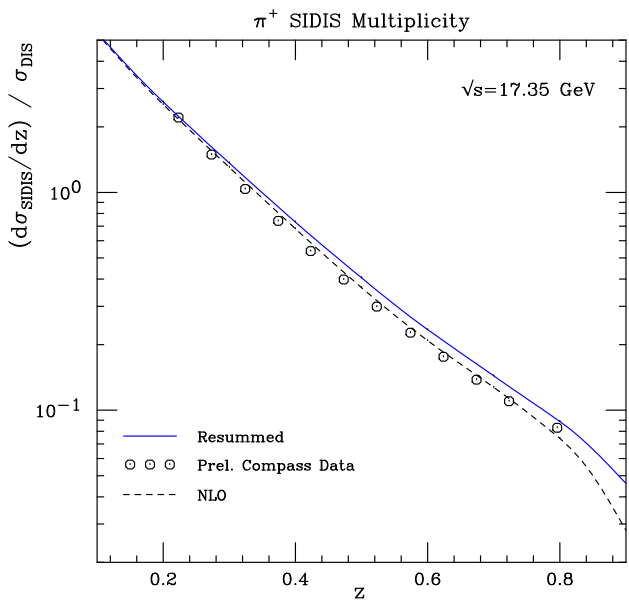}
\vspace*{-1.5cm}
\caption{\label{fig6}\sf NLO and NLL resummed SIDIS multiplicities for $\pi^+$. The results are
compared to the preliminary COMPASS data~\cite{compass}. The uncertainties of the data
are smaller than the symbol size used in the plot.}
\end{figure}

\begin{figure}[t]
\vspace*{-9mm}
\hspace*{-3.8cm}
\includegraphics[width=0.55\textwidth,angle=90]{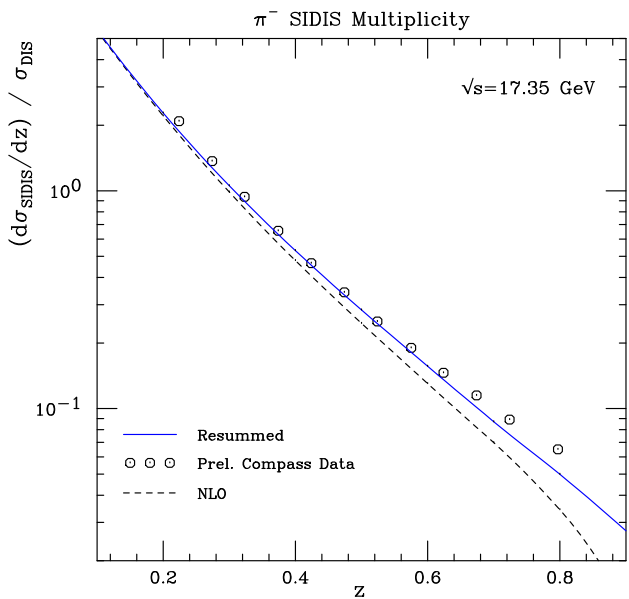}
\vspace*{-1.5cm}
\caption{\label{fig7}\sf Same as Fig.~\ref{fig6}, but for $\pi^-$.}
\end{figure}

Preliminary precise data from COMPASS for charged-pion and kaon multiplicities are available for a wide range 
of kinematics~\cite{compass}. The full data set, which evidently has the best statistics, covers the range used 
above for Figs.~\ref{fig4} and~\ref{fig5}. Figures~\ref{fig6} and~\ref{fig7} show comparisons of our
NLO and NLL resummed calculations for charged pions to the COMPASS data. As one can see,
resummation leads to a moderate, but significant, enhancement of the multiplicities. It is interesting
to note that such an enhancement is in fact preferred by the $\pi^-$ data. However, we do not
assign much importance to this observation. The fragmentation functions are presently 
still not very well determined and a different set (or a new fit) might well describe the data also at NLO. 
Our main point is that inclusion of resummation effects in an analysis of fragmentation functions 
could make a significant difference for the extracted functions.
%%%%%%%%%%%%%%%%%%%%%%%%%%%%%%%%%%%%%%%%%%%%%%%%
%%%%%%%%%%%%%%%%%%%%%%%%%%%%%%%%%%%%%%%%%%%%%%%%
%%%%%%%%%%%%%%%%%%%%%%%%%%%%%%%%%%%%%%%%%%%%%%%%
%%%%%%%%%%%%%%%%%%%%%%%%%%%%%%%%%%%%%%%%%%%%%%%%

\begin{figure}[t]
\vspace*{-9mm}
\hspace*{-3.8cm}
\includegraphics[width=0.55\textwidth,angle=90]{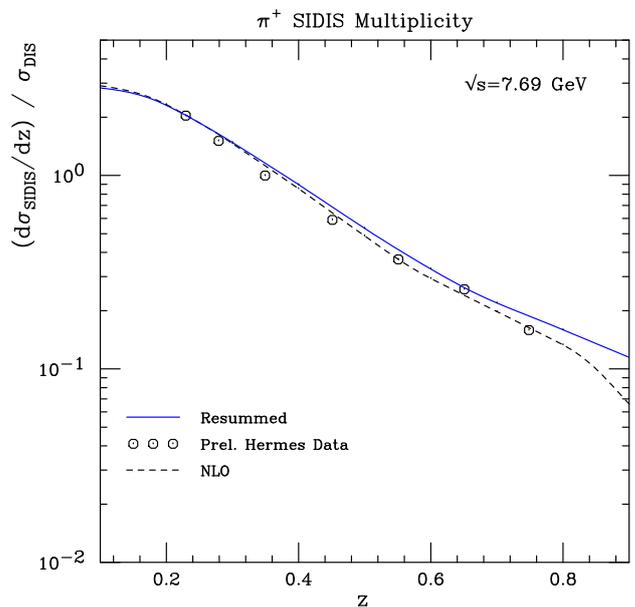}
\vspace*{-1.5cm}
\caption{\label{fig8}\sf Same as Fig.~\ref{fig6}, but for HERMES kinematics. The
preliminary data are from~\cite{hermes}.  The uncertainties of the data
are smaller than the symbol size used in the plot.}
\end{figure}

\begin{figure}[h]
\vspace*{-6mm}
\hspace*{-3.8cm}
\includegraphics[width=0.55\textwidth,angle=90]{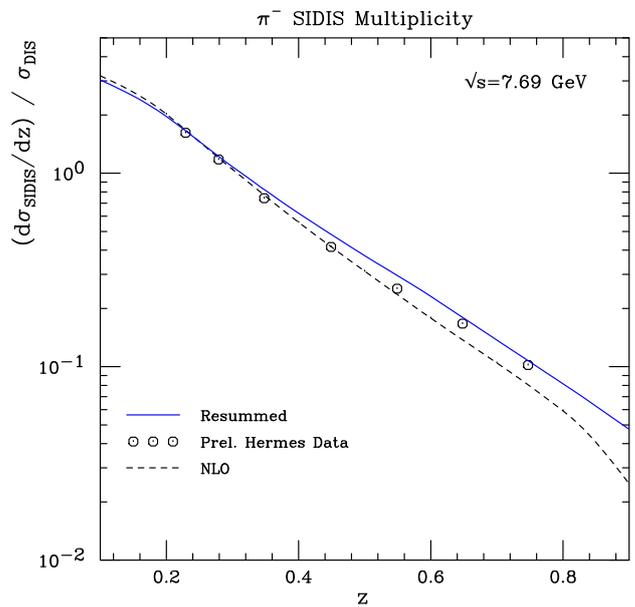}
\vspace*{-1.5cm}
\caption{\label{fig9}\sf Same as Fig.~\ref{fig8}, but for $\pi^-$.}
\end{figure}

Figures~\ref{fig8} and~\ref{fig9} show similar comparisons to the HERMES preliminary
data~\cite{hermes}. For this data set, the center-of-mass energy is $\sqrt{s}\approx$17.4 GeV. 
The kinematic cuts employed by HERMES are $0.023<x<0.6$, $0.1<y<0.85$, $Q^2>1$ GeV${}^2$ 
and $W^2=10$ GeV${}^2$. The results are qualitatively similar to those shown for COMPASS
kinematics. 
%%%%%%%%%%%%%%%%%%%%%%%%%%%%%%%%%%%%%%%%%%%%%%%%
%%%%%%%%%%%%%%%%%%%%%%%%%%%%%%%%%%%%%%%%%%%%%%%%
%%%%%%%%%%%%%%%%%%%%%%%%%%%%%%%%%%%%%%%%%%%%%%%%
%%%%%%%%%%%%%%%%%%%%%%%%%%%%%%%%%%%%%%%%%%%%%%%%

\subsection{Results for single-inclusive $e^+e^-$ annihilation}

\begin{figure}[t]
\vspace*{-6mm}
\hspace*{-3.8cm}
\includegraphics[width=0.55\textwidth,angle=90]{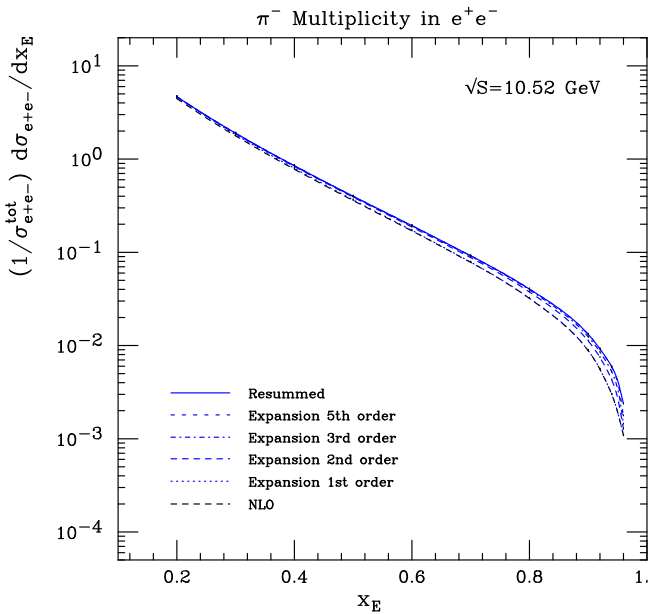}
\vspace*{-1.5cm}
\caption{\label{fig10}\sf $\pi^-$ multiplicity in electron-positron annihilation at $\sqrt{s}=10.52$~GeV. 
The two lowest lines show the NLO result and the (unmatched) first-order expansion of the resummed one, which
are practically indistinguishable. The other lines show matched higher-order expansions of the resummed
multiplicity, and the resummed result itself (solid line).}
\end{figure}

\begin{figure}[h]
\vspace*{-8mm}
\hspace*{-4.1cm}
\includegraphics[width=0.57\textwidth,angle=90]{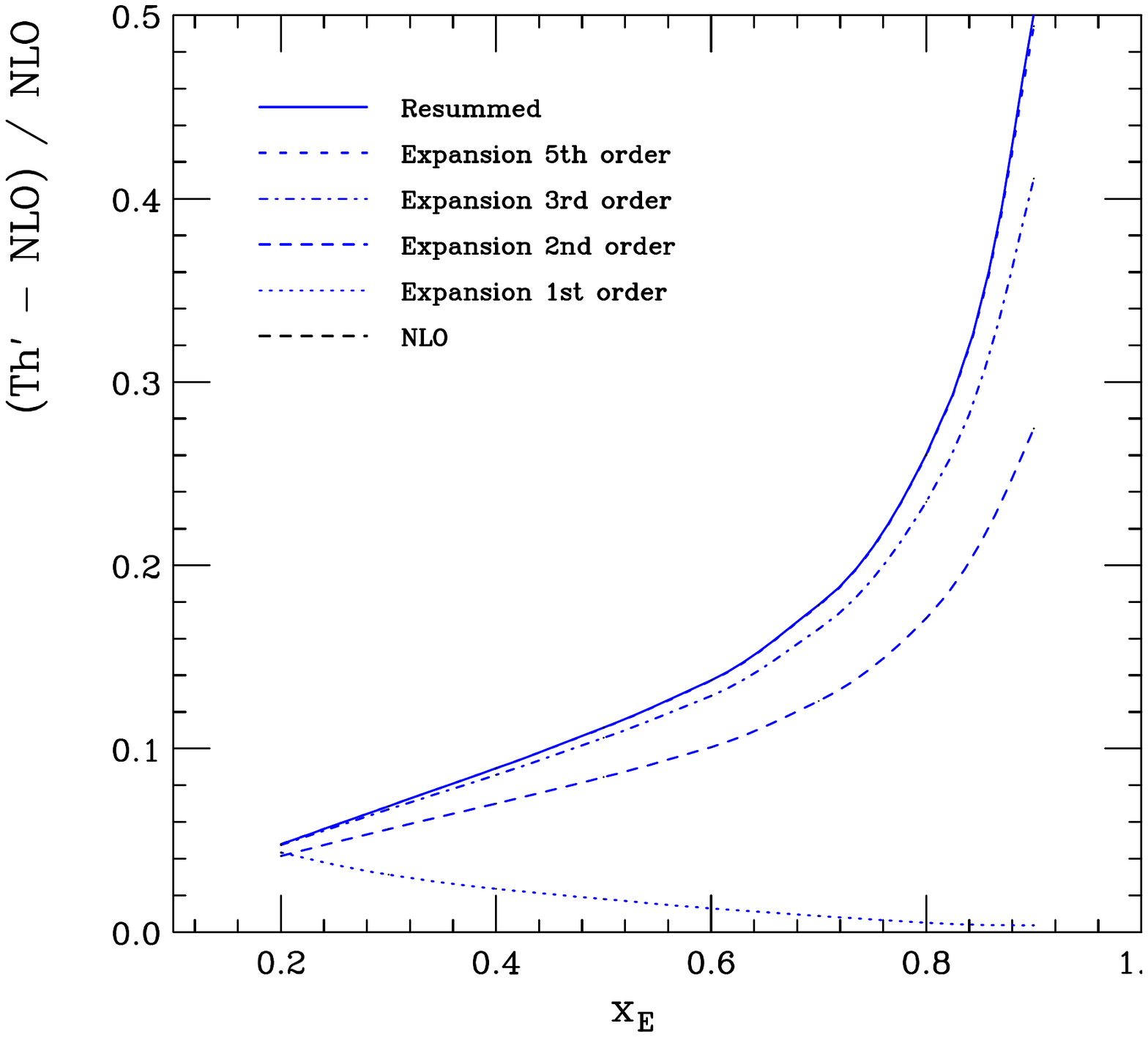}
\vspace*{-1.5cm}
\caption{\label{fig11}\sf Ratios $({\mathrm{Th' - NLO}})/{\mathrm{NLO}}$
corresponding to the various curves shown in Fig.~\ref{fig10}.}
\end{figure}

Figure~\ref{fig10} presents our results for the $\pi^-$ multiplicity in $e^+e^-$ annihilation
at $\sqrt{s}=10.52$~GeV and for $-1<\cos\theta<1$, as appropriate for comparison to the 
forthcoming BELLE data~\cite{belle}. The $\pi^+$ multiplicity is identical thanks to charge conjugation 
symmetry. We have chosen the factorization and renormalization scales as $\sqrt{s}$.
As for SIDIS, we show NLO and NLL resummed results, along with various fixed-order 
expansions of the resummed multiplicity. One first of all observes the excellent agreement between 
the NLO result and the first-order expansion of the resummed one. This clearly demonstrates that
the threshold regime strongly dominates for the BELLE kinematics. We can therefore
be confident that also the resummed result reliably captures the important higher-order terms. 
Resummation leads to a significant enhancement of the $\pi^-$ multiplicity. This enhancement
becomes particularly strong at high $x_E$, but is present also at moderate values. Figure~\ref{fig11}
shows the corresponding ratios $({\mathrm{Th' - NLO}})/{\mathrm{NLO}}$, where
${\mathrm{Th'}}$ denotes any of the higher-order multiplicities generated by 
resummation and shown in Fig.~\ref{fig10}. Since the precision 
of the preliminary BELLE data is typically much better than 10\%, it will be important to 
include the enhancements we find in future global analyses of fragmentation functions, 
similar to what has been done in the past in Ref.~\cite{akk}.
%%%%%%%%%%%%%%%%%%%%%%%%%%%%%%%%%%%%%%%%%%%%%%%%
%%%%%%%%%%%%%%%%%%%%%%%%%%%%%%%%%%%%%%%%%%%%%%%%
%%%%%%%%%%%%%%%%%%%%%%%%%%%%%%%%%%%%%%%%%%%%%%%%
%%%%%%%%%%%%%%%%%%%%%%%%%%%%%%%%%%%%%%%%%%%%%%%%

%%%%%%%%%%%%%%%%%%%%%%%%%%%%%%%%%%%%%%%%%%%%%
\section{Conclusions \label{sum}}
%%%%%%%%%%%%%%%%%%%%%%%%%%%%%%%%%%%%%%%%%%%%%

We have derived threshold-resummed expressions for the coefficient functions for single-inclusive
hadron production in semi-inclusive lepton scattering and $e^+e^-$ annihilation. We have presented
phenomenological results for pion multiplicities for these processes in the kinematic regimes
presently accessed by the COMPASS, HERMES and BELLE experiments. We have found that 
resummation leads to modest but significant enhancements of the multiplicities. The recent
preliminary SIDIS data turn out to be overall better described when resummation effects
are included in the calculation, at least for the DSS set of fragmentation functions that we have used. 
However, we do not ascribe much significance to this point as the fragmentation functions 
are still rather poorly constrained so that a new NLO fit to the new data would likely also 
work well~\cite{strat}. Our main point is that, given the good accuracy of the new preliminary
SIDIS and BELLE data,  it will be crucial to include resummation effects for both processes
in the next generation of analyses of fragmentation functions. In fact, because of the
enhancements from threshold resummation, the extracted fragmentation functions would
be expected to be softer or smaller at high $z$ than functions extracted purely on the basis
of an NLO framework. This may well have important ramifications for the QCD predictions
obtained for other processes sensitive to fragmentation functions. For instance, it has recently been
observed~\cite{strat,dent} that the ALICE data~\cite{Abelev:2012cn} for neutral-pion production 
at 7~TeV are well below the theoretical NLO expectations. One may speculate if this is due in part to
fragmentation functions that are too large at high $z$. 

We note that at very high $z$ or $x_E$, besides the logarithmic perturbative corrections also
nonperturbative power corrections will ultimately become relevant and will need to be analyzed theoretically. 
Resummation offers ways to address these contributions (see, for example~\cite{Sterman:2006hu,power}, 
and references therein). Based on the ideas presented there, we do not think that power corrections play an 
overwhelming role in the presently accessible kinematic regime in SIDIS. As we saw in Eq.~(\ref{resummed4}), 
the logarithmic contributions to SIDIS come from the region $k_\perp\gtrsim Q/\sqrt{\bar{N}\bar{M}}$. 
In $x$-$z$-space this scale roughly corresponds to $Q\sqrt{(1-z)(1-x)}$. We have checked that
for COMPASS kinematics even at $z=0.9$ the average value of this scale is significantly larger
than 1~GeV, implying that perturbation theory should still provide a reliable answer here. This
issue obviously deserves a more detailed investigation in the future, in particular also for the case
of $e^+e^-$ annihilation. 

We stress that our study may be extended in several ways. First, as mentioned in the 
Introduction, resummation for $e^+e^-\to hX$ could be carried out at next-to-next-to-leading logarithm 
and even beyond, thanks to~\cite{Blum,Vogt}. The high precision of the BELLE
data may well warrant a future study along these lines. Resummation for SIDIS could probably
also be extended to next-to-next-to-leading logarithmic accuracy with some moderate further
developments. This may become a very worthwhile task in the future when high-precision 
SIDIS data will become available from measurements at the Jefferson Laboratory after the CEBAF upgrade 
to 12~GeV beam energy~\cite{Burkert:2012rh}. Finally, one may readily adapt the resummation 
framework to the case of polarized SIDIS, which serves as an important probe of the spin
structure of the nucleon. Present polarized-SIDIS measurements have the same kinematics
as those considered in this paper.

%%%%%%%%%%%%%%%%%%%%%%%%%%%%%%%
\section{Acknowledgments}
%%%%%%%%%%%%%%%%%%%%%%%%%%%%%%%
      
We thank M. Traini for helpful discussions and M. Leitgab for useful communications
on the BELLE measurements.

\section*{Appendix}
In this Appendix we collect the NLO expressions for the spin-averaged partonic SIDIS
cross sections. At NLO, we have to consider the processes $\gamma^* q\to q X$, 
$\gamma^* q\to g X$, and  $\gamma^* g\to q X$. For their contributions to the 
structure function ${\cal F}_T^h$ we have in the $\overline{\mathrm{MS}}$ 
scheme~\cite{altarelli,Nason:1993xx,fupe,graudenz,deFlorian:1997zj,roth}: 
\begin{eqnarray}
\label{sidiseq8}
\nonumber
C_{qq}^{T,(1)}(\hat{x},\hat{z}) &=& e_q^2 C_F
\Bigg[ -8\delta(1-\hat{x})\delta(1-\hat{z})
\\[2mm]
\nonumber
&&\hspace*{-1.6cm}+
\delta(1-\hat{x}) \left[ \tilde{P}_{qq}(\hat{z}) \ln\frac{Q^2}{\mu_F^2} +
 L_1(\hat{z})+L_2(\hat{z})+(1-\hat{z})\right]  \\[2mm]
\nonumber
&&\hspace*{-1.6cm}
+\delta(1-\hat{z}) \left[ \tilde{P}_{qq}(\hat{x}) \ln\frac{Q^2}{\mu_F^2} +
L_1(\hat{x})-L_2(\hat{x})+(1-\hat{x})\right]\\[2mm]
&& \hspace*{-1.6cm}+\frac{2}{(1-\hat{x})_+(1-\hat{z})_+}
- \frac{1+\hat{z}}{(1-\hat{x})_+}-
\frac{1+\hat{x}}{(1-\hat{z})_+}\nn\\[2mm]
&&\hspace*{-1.6cm}+2(1+\hat{x}\hat{z}) \Bigg],
\end{eqnarray}
\begin{eqnarray}
\label{sidiseq9}
\nonumber
C_{gq}^{T,(1)}(\hat{x},\hat{z}) &=&  e_q^2 C_F
\Bigg[ \tilde{P}_{gq}(\hat{z}) \left(\delta (1-\hat{x}) \ln\left(
\frac{Q^2}{\mu_F^2}\hat{z}(1-\hat{z})\right)\right.\\[2mm]
&&\hspace*{-1.6cm}\left.+\frac{1}{(1-\hat{x})_+}\right)+
\hat{z} \delta(1-\hat{x})+2(1+\hat{x}-\hat{x}\hat{z})-\frac{1+\hat{x}}{\hat{z}}\Bigg],\nn\\[2mm]
\end{eqnarray}
\begin{eqnarray}
\label{sidiseq10}
C_{qg}^{T,(1)}(\hat{x},\hat{z}) &=& e_q^2 T_R 
\Bigg[ \delta (1-\hat{z}) \left[\tilde{P}_{qg}(\hat{x})
\ln\left(\frac{Q^2}{\mu_F^2} \frac{1-\hat{x}}{\hat{x}}\right) \right.\nn\\[2mm]
&&\hspace*{-2cm}+2\hat{x}(1-\hat{x})\Big] 
+ \tilde{P}_{qg}(\hat{x}) \left\{ \frac{1}{(1-\hat{z})_+}+\frac{1}{\hat{z}}-2\right\} \Bigg],
\end{eqnarray}
where $e_q$ is the quark's fractional charge, $C_F=4/3$, $T_R=1/2$, 
\begin{eqnarray}
\nonumber
\label{sidiseq6}
&&\tilde{P}_{qq} (\xi)= \frac{1+\xi^2}{(1-\xi)_+} + \frac{3}{2} 
\delta (1-\xi),\nonumber\\[2mm]
&&\tilde{P}_{gq}(\xi)=\frac{1+(1-\xi)^2}{\xi},\nonumber\\[2mm]
&&\tilde{P}_{qg}(\xi)= \xi^2+(1-\xi)^2, \nn\\[2mm]
&&L_1(\xi)=(1+\xi^2)\left(\frac{\ln (1-\xi)}{1-\xi}\right)_+,\nn\\[2mm]
&&L_2(\xi)=\frac{1+\xi^2}{1-\xi}\ln \xi ,
\end{eqnarray}
and the ``+'' - distributions are defined as follows:
\begin{eqnarray}
\label{sidiseq25}
\int_0^1 d\xi f(\xi)[g(\xi)]_+ &\equiv&
\int_0^1d\xi \left(f(\xi)-f(1)\right) g(\xi),\nn\\[2mm]
&&\hspace*{-3.4cm}\int_0^1 d\hat{x}\int_0^1 d\hat{z} 
\frac{f(\hat{x},\hat{z})}{(1-\hat{x})_+ (1-\hat{z})_+} \nn\\[2mm]
&&\hspace*{-3cm}\equiv
\int_0^1 d\hat{x}\int_0^1  d\hat{z}\, 
\frac{f(\hat{x},\hat{z})-f(1,\hat{z})-f(\hat{x},1)+f(1,1)}{(1-\hat{x})(1-\hat{z})}.\nn\\[2mm]
\end{eqnarray}
Note that we have given expressions~(\ref{sidiseq8})-(\ref{sidiseq10}) for an arbitrary
factorization scale $\mu_F$, keeping however the scales the same for the initial and 
the final state. For the longitudinal structure function $F^h_L$:
\begin{eqnarray}
C_{qq}^{L,(1)}(\hat{x},\hat{z}) &=& 4 e_q^2 C_F \hat{x} \hat{z} ,\nn\\[2mm]
C_{gq}^{L,(1)}(\hat{x},\hat{z}) &=& 4 e_q^2 C_F \hat{x} (1-\hat{z}),\nn \\[2mm]
C_{qg}^{L,(1)}(\hat{x},\hat{z}) &=& 8 e_q^2 T_R \hat{x}(1-\hat{x}) .
\end{eqnarray}

In Mellin-moment space, the NLO results become~\cite{Stratmann:2001pb}
\begin{eqnarray} 
\label{cqq}
\tilde{C}_{qq}^{T,(1)}(N,M)&=&e_q^2 C_F \Bigg[
-8-\frac{1}{M^2} +\frac{2}{(M+1)^2}+\frac{1}{N^2}\nn\\[2mm]
&&\hspace*{-1.6cm}
+\frac{(1+M+N)^2+1}{M (M+1)N(N+1)}+3 S_2(M)  - S_2(N)\nn\\[2mm] 
&&\hspace*{-1.6cm}+ \left[ S_1(M) + S_1(N) \right] \Bigg\{ 
S_1(M) + S_1(N)\nn\\[2mm]
&&\hspace*{-1.6cm}\left.-\frac{1}{M(M+1)}-\frac{1}{N(N+1)}\right\} \nonumber \\[2mm]
&&\hspace*{-1.6cm}
+ \left[\frac{1}{N(N+1)}+\frac{3}{2}-2 S_1(N)\right] \ln\left(\frac{Q^2}{\mu_F^2}\right)
 \nonumber \\ [2mm] 
&&\hspace*{-1.6cm}
+ \left[\frac{1}{M(M+1)}+\frac{3}{2}-2 S_1(M)\right] \ln\left(\frac{Q^2}{\mu_F^2}\right)
\Bigg]  , 
\end{eqnarray}
\begin{eqnarray}
\label{cgq}
\tilde{C}_{gq}^{T,(1)}(N,M)&=& e_q^2 C_F
 \Bigg[
\frac{2-2 M-9 M^2+M^3-M^4+M^5}{M^2 (M-1)^2(M+1)^2}\nn\\[2mm]
&&\hspace*{-1.6cm}+\frac{2M}{N (M+1)(M-1)}-\frac{2-M+M^2}{M (M+1)(M-1)(N+1)}
\nn\\[2mm]
&&\hspace*{-1.6cm}-\frac{2+M+M^2}{M (M+1)(M-1)} 
\left[ S_1(M) + S_1(N) \right]  \nonumber \\ [2mm]
&&\hspace*{-1.6cm}+ \frac{2+M+M^2}{M(M+1)(M-1)} \ln\left(\frac{Q^2}{\mu_F^2}\right)
\Bigg] , 
\end{eqnarray}
\begin{eqnarray}
\tilde{C}_{qg}^{T,(1)} (N,M)&=& e_q^2 T_R
\Bigg[\frac{2+N+N^2}{N(N+1)(N+2)} \Bigg(
\frac{1}{M-1}-\frac{1}{M} \nn\\[2mm]
&&\hspace*{-1.6cm}
-S_1(M)-S_1(N)+ \ln\left(\frac{Q^2}{\mu_F^2}\right)
\Bigg) +\frac{1}{N^2} \Bigg],
\end{eqnarray}
where
\begin{equation}
S_i(N) \equiv \sum_{j=1}^N \frac{1}{j^i} \; .
\end{equation}
Note that at large $N$ we have
\beq
S_1(N)=\ln\bar{N}+{\cal O}(1/N),\;\;S_2(N)=\frac{\pi^2}{6}
+{\cal O}(1/N),
\eeq
where $\bar{N}=N{\mathrm{e}}^{\gamma_E}$.
Furthermore, for the longitudinal structure function,
\begin{eqnarray}
\tilde{C}_{qq}^{L,(1)}(N,M)&=& e_q^2C_F \frac{4}{(M+1)(N+1)}, \\ [1.5mm]
\tilde{C}_{gq}^{L,(1)}(N,M) &=& e_q^2C_F \frac{4}{M(M+1)(N+1)}, \\ [1.5mm]
\tilde{C}_{qg}^{L,(1)}(N,M) &=& e_q^2T_R\frac{8}{M(N+1)(N+2)}.
\end{eqnarray}
Here we have corrected a mistake in $\tilde{C}_{qg}^{L,(1)}(N,M)$ in Ref.~\cite{Stratmann:2001pb}.

\newpage
%
%%%%%%%%%%%%%%%%%%%%%%%%%%%

\end{document}